\documentclass{iaus}
\usepackage{graphics}

  \checkfont{eurm10}
  \iffontfound
    \IfFileExists{upmath.sty}
      {\typeout{^^JFound AMS Euler Roman fonts on the system,
                   using the 'upmath' package.^^J}%
       \usepackage{upmath}}
      {\typeout{^^JFound AMS Euler Roman fonts on the system, but you
                   dont seem to have the}%
       \typeout{'upmath' package installed. iaus.cls can take advantage
                 of these fonts,^^Jif you use 'upmath' package.^^J}%
      }
  \else
  \fi

  \checkfont{msam10}
  \iffontfound
    \IfFileExists{amssymb.sty}
      {\typeout{^^JFound AMS Symbol fonts on the system, using the
                'amssymb' package.^^J}%
       \usepackage{amssymb}%
         
       \let\ge=\geqslant  
      }{}
  \fi

  \IfFileExists{amsbsy.sty}
    {\typeout{^^JFound the 'amsbsy' package on the system, using it.^^J}%
     \usepackage{amsbsy}}
    {\providecommand\boldsymbol[1]{\mbox{\boldmath $##1$}}}

\newsavebox{\astrutbox}
\sbox{\astrutbox}{\rule[-5pt]{0pt}{20pt}}

\title[Outskirts of Galaxy Clusters: intense life in the suburbs]
      {The Luminosity Function of Galaxies in the Hercules Cluster}

\author[R. S\' anchez-Janssen {\it et al.\/}]%
{R. S\' anchez-Janssen$^1$%
, J. Iglesias-P\'aramo$^2$
, C. Mu\~noz-Tu\~n\'on$^1$, \break
J. A. L. Aguerri$^1$
\and J. M. V\'\i lchez$^3$}

\affiliation{$^1$Instituto de Astrof\'\i sica de Canarias, Spain; email: ruben@ll.iac.es\\[\affilskip]
$^2$Laboratoire d'Astrophysique de Marseille, France\\[\affilskip]
$^3$Instituto de Astrof\'\i sica de Andaluc\'\i a, CSIC, Spain}

\pubyear{2004}
\volume{195}
\pagerange{1--8}
\date{?? and in revised form ??}
\jname{Outskirts of Galaxy Clusters: intense life in the suburbs}
\editors{A. Diaferio, ed.}
\begin{document}

\maketitle

\begin{abstract}
 We have imaged $\sim$ 1 deg$^{2}$ in the V-band in the direction of the Hercules Cluster (Abell 2151). The data are used to compute for the first time the luminosity function (LF) of galaxies in the cluster down to the dwarf regime (M$_{lim}$ $\sim$ -13.85 ). The global LF is well described by a Schechter function (\cite{schechter76}) with best-fit parameters $\alpha$ = -1.30 $\pm$ 0.06 and M$_V$$^*$ = -21.25 $\pm$ 0.25. The radial dependence of the LF has also been studied, finding that it turns out to be almost constant within the errors even further away than the virial radius. Given the presence of significant substructure within the cluster, we have analized the LFs in different regions. While two of the subclusters present LFs consistent with each other and with the global one, the southernmost one exhibits a somewhat steeper faint-end slope.
\end{abstract}

\firstsection 
\section{Introduction}

The luminosity function of galaxies (LF), that is, the probability density of galaxies in a certain population having a given luminosity, is one of the key observable quantities for galaxy evolution theories. The comparison between the LF of galaxies in clusters and in the field should clarify the role played by the environment in regulating galaxy evolution. Recent results by large surveys (\cite{depropris03}) point towards the composite LF of cluster galaxies having a steeper faint-end slope than field galaxies, revealing that this influence indeed exists. A2151 is a nearby (z = 0.0367), irregular and spiral-rich cluster ($\sim$ 50\%). There are strong evidences (from optical and X-ray studies) suggesting that the cluster is still on the process of collapse: the lack of hydrogen deficiency in the spiral population (\cite{GH85}), the bumpy distribution of the hot intracluster gas and its low X-ray flux (\cite{HS96}), and the presence of at least three distinct subclusters (\cite{BDB95}, BDB hereafter) point towards A2151 being a young and relatively unevolved cluster, thus making it an excellent target for the study of the LF and stablishing comparisons with more evolved systems.

\section{Observations}\label{sec:observations}

The observations were carried out with the WFC at the 2.5m INT. We have imaged a mosaic of five overlapping pointings covering an area of about $1^{\circ}\times1^{\circ} $ across the cluster (Fig.~\ref{fig:figure} \textit{left}). The calibration was achieved using aperture magnitudes of bright galaxies taken from \cite{TKK95}. We have also observed the nearby SA107 control field for purposes of background galaxy counts decontamination. The V magnitudes were corrected for Galactic extinction using the \cite{schlegel98} dust map and the \cite{CCM89} extinction curve.

\section{Source extraction and photometry}\label{sex}

The identification and extraction of sources was carried out using SExtractor (\cite{BA96}). The separation between stars and galaxies was performed based on SExtractor's stellarity index (S/G), which assigns a numerical value close to 1 if the object is a star and values closer to 0 if it is a galaxy-like object. Pitifully, the S/G classifier breaks down at the lower magnitude end, complicating the identification. In order to test the reliability of the index we simulated artificial images of isolated gaussian-PSF stars and recovered them with the same selection criteria as used for the Hercules fields. FWHMs ranged the values of our frames. We obtained values of S/G $\ge$ 0.85 for star-like objects and with accurate photometry down to V $\sim$ 22.

\section{Sample completeness}\label{sec:simulations}

One of the key problems in the study of LFs is the knowledge of detection limits of the data. We have carried out Monte Carlo simulations of artificial galaxy images, and, as done with stars, recovered them using SExtractor. We generated galaxies with four different morphologies: three exponential profiles with r$_{eff}$ = 0.75, 1.5 and 3 kpc and one S\' ersic profile with r$_{eff}$ = 0.5 kpc, each in the magnitude range 17 $<$ V $<$ 23 (-18.85 $<$ M$_V$ $<$ -12.85 at the distance of A2151 and assuming a value of H$_{0}$ = 75 km s$^{-1}$ Mpc$^{-1}$). Those are typical parameters of dE and dSph galaxies in nearby clusters (\cite{BJ98}) which are supposed to dominate the faint end of the LF. Results of our simulations show that, as expected, galaxies with higher effective radii are hardly detected, while recovered magnitude differences are always below 0.3 mag, which do not play an important role in our subsequent analysis since for the LF we are using bins of 1 mag width. If we make the simple hypothesis that all these galaxies occur with the same probability in the cluster, our sample is complete at the 80\% down to V = 22.

\section{The Luminosity Functions}\label{sec:LF}

\begin{flushleft}
\begin{figure*}

\begin{tabular}{cc}
  \scalebox{0.40}{\includegraphics{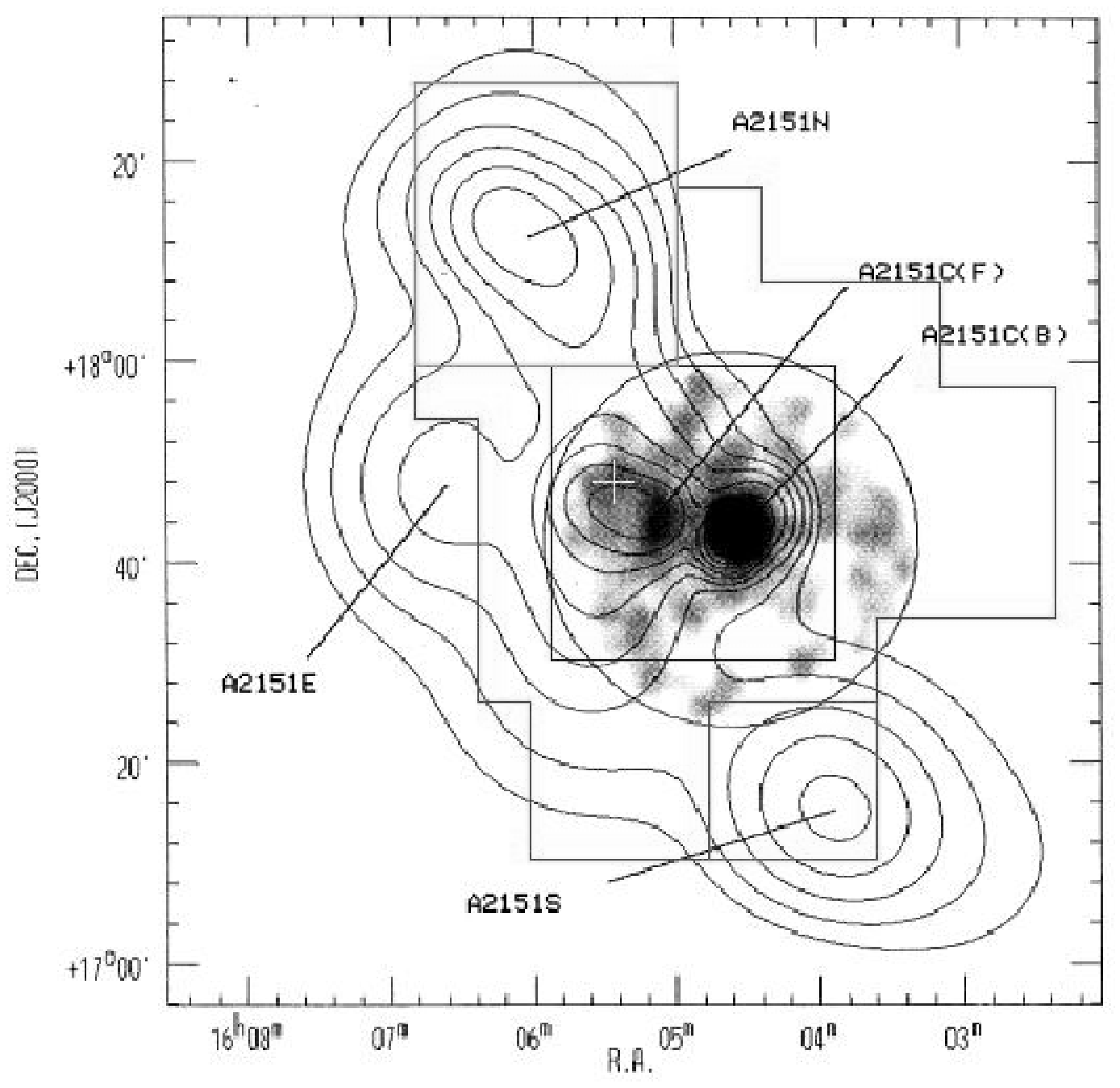}} & \scalebox{0.50}{\includegraphics{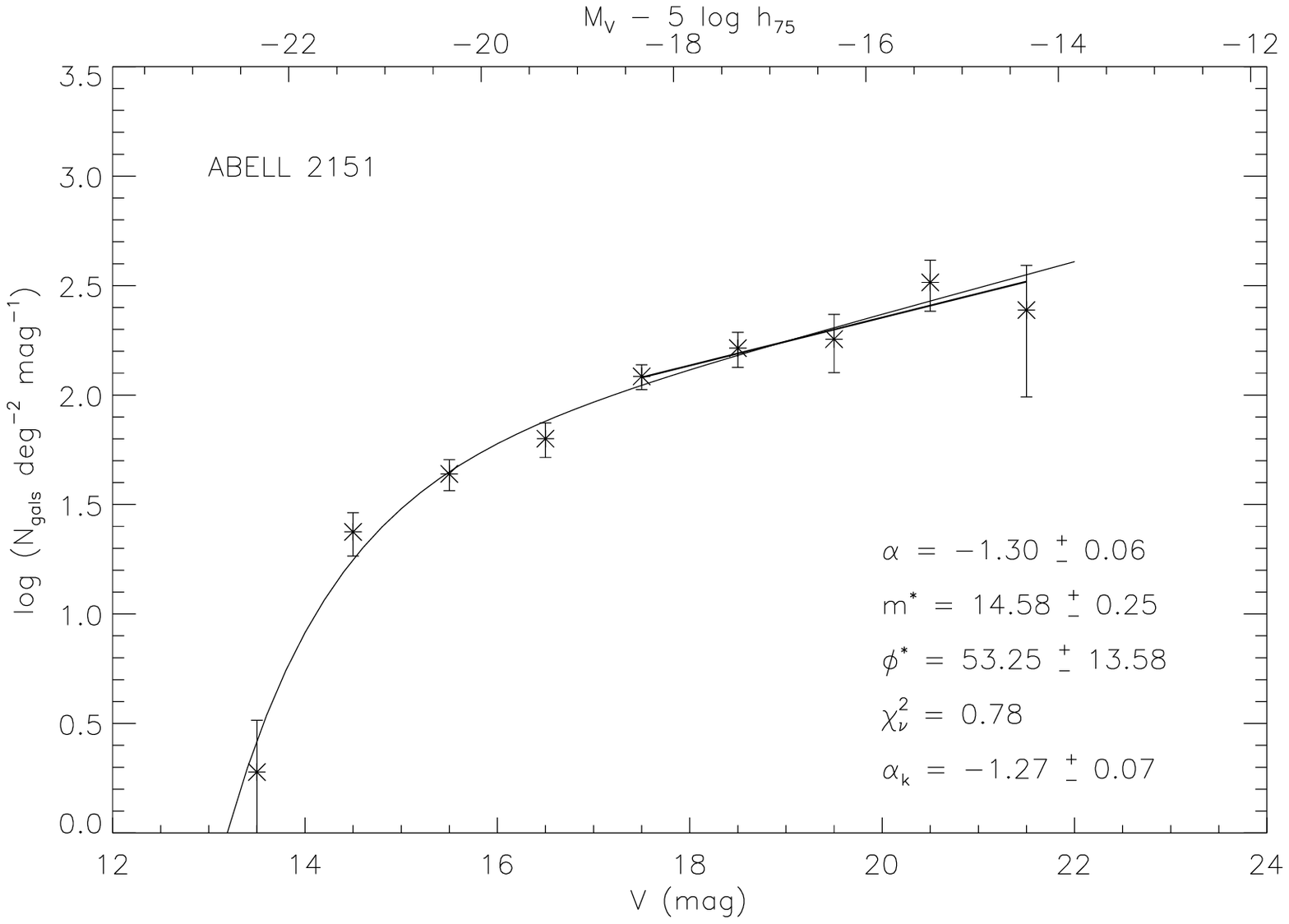}}
\end{tabular}

  \caption{
    (\textit{Left}) Surveyed area plotted over galaxy density contours and X-ray emission (from \cite{HS96}). Small squares show the areas where subclusters' LFs have been computed.
    (\textit{Right}) The global LF of A2151 and its best-fit Schechter function. Error bars also include non-Poissonian fluctuations.}\label{fig:figure}

\end{figure*}
\end{flushleft}

Once we had the number counts and effective areas for both the cluster and control field samples, we computed the Hercules Cluster LF as the statistical difference between these counts. In Fig.~\ref{fig:figure} (\textit{right}) we plot the cluster LF down to V = 22 (M$_{V} \sim$ -13.85), where the error bars include the calibration uncertainties and both Poissonian and non-Poissonian fluctuations of counts, the latter being due to the cosmic variance (\cite{huang97}). We fitted a Schechter function to the data, finding that it is a good representation of the LF ($\chi _{\nu} ^{2}$ = 0.78). Best-fit parameters are $\alpha$ = -1.30$\pm$0.06, M$_{V}$$^{*}$ = -21.25$\pm$0.25 and  $\phi^{*}$ = 53.25$\pm$13.58. \cite{lugger86} computed the R-band LF of A2151 in a region 4.6 Mpc a side, but only down to R $\sim$ -19.8 (six magnitudes brighter than our limit). She derived a value of $\alpha = -1.26\pm0.13$, similar to ours despite of the differences between both works.\\

We have also investigated the radial dependence of Schechter parameters as we move outwards the cluster. We divided our area in four different annuli from inside the core radius to outside the virial radius, and computed the LFs in them. The faint-end slope turned out to be almost constant within the errors. \\

A2151 shows substantial subclustering with at least three distinct subclusters identified by BDB. In our mapping we cover two of them (A2151N and A2151C), while A2151E is out of our survey (see ~\ref{fig:figure} \textit{left}). However, a fourth overdensity (A2151S) seems to exist in the southernmost part of the cluster, though it was not identified as a distinct subcluster in the BDB analysis. We have computed the LFs of A2151N, A2151C and (partly) A2151S in the areas shown in Fig.~\ref{fig:figure} (\textit{left}) down to V=21. We find that, within the errors,  the two identified subclusters have slopes consistent with each other and with the global one ($\alpha$ = -1.33$\pm$0.06 and $\alpha$ = -1.17$\pm$0.16 for A2151C and A2151N respectively) . However, for A2151S we surprisingly found a somewhat steeper slope ($\alpha$ = -1.64$\pm$0.10). A detailed inspection of the LF suggests that this increase in the dwarf to giant ratio (D/G) is mainly due to a significant population of faint galaxies. If this increment is not due to a contamination by background galaxies in this region, it would imply the presence of a large population of dwarf galaxies in this southern subcluster.

\begin{acknowledgments}
RSJ acknowledges financial support from the \emph{Academia Canaria de Ciencias} and thanks the other authors for their continuous support.
\end{acknowledgments}


\begin{thebibliography}{}







  \bibitem[Bertin \& Arnouts 1996]{BA96}
     {Bertin, E. \& Arnouts, S.} 1996
     A\&A \textbf{117}, 393.

   \bibitem[Binggeli \& Jerjen 1998]{BJ98}
     {Binggeli, B. \& Jerjen, H.} 1998
     A\&A \textbf{333}, 17.

  \bibitem[Bird, Davis \& Beers 1995]{BDB95}
     {Bird, C.M., Davis, D.S.  \& Beers, T.C.} 1995
     AJ \textbf{109}, 920.

   \bibitem[Cardelli, Clayton \& Mathis (1989)]{CCM89}
     {Cardelli, J.A., Clayton, G.C. \& Mathis, J.S.} 1989
     A\&A \textbf{345}, 245.

   \bibitem[De Propris et al. 2003]{depropris03}
     {De Propris et al.} 2003
     MNRAS \textbf{342}, 725.

   \bibitem[Giovanelli \& Haynes 1985]{GH85}
     {Giovanelli, R. \& Haynes, M.P.} 1985
     ApJ \textbf{292}, 404.

   \bibitem[Huang \& Sarazin 1996]{HS96}
     {Huang, Z. \& Sarazin, C.L. } 1996
     ApJ \textbf{461}, 622.

   \bibitem[Huang et al. 1997]{huang97}
     {Huang, J.-S, Cowie, L.L., Gardner, J.P., Hu, E.M., Songaila, A. \& Wainscoat, R.J.} 1997
     ApJ \textbf{476}, 12.

   \bibitem[Lugger (1986)]{lugger86}
     {Lugger, P.M.} 1986
     ApJ \textbf{303}, 535.

   \bibitem[Schechter 1976]{schechter76}
     {Schechter, P.} 1976
     ApJ \textbf{203}, 297.

   \bibitem[Schlegel, Finkbeiner \& Davis (1998)]{schlegel98}
     {Schlegel, D.J., Finkbeiner, D.P. \& Davis, M.} 1998
     ApJ \textbf{500}, 525.

   \bibitem[Takamiya, Kron \& Kron (1995)]{TKK95}
     {Takamiya, M., Kron, R.G. \& Kron, G.E.} 1995
     AJ \textbf{110}, 1083.



\end{thebibliography}
\end{document}